# Multiscale stress dynamics in sheared liquid foams revealed by tomo-rheoscopy

Florian Schott[a], Benjamin Dollet[b], Stéphane Santucci[c], Christian Matthias Schlepütz[d], Cyrille Claudet[e], Stefan Gstöhl[d], Christophe Raufaste[e,f], and Rajmund Mokso[g,a]

October 21, 2024

**Rheology aims at quantifying the response of materials to mechanical forcing. However, standard rheometers provide only global macroscopic quantities, such as viscoelastic moduli. They fail to capture the heterogeneous flow of soft amorphous materials at the mesoscopic scale, arising from the rearrangements of the microstructural elements, that must be accounted for to build predictive models. To address this experimental challenge, we have combined shear rheometry and time-resolved X-ray micro-tomography on 3D liquid foams used as model glassy materials, yielding a unique access to the stresses and contact network topology at the bubble scale. We reveal a universal scaling behavior of the local stress build-up and relaxation associated with topological modifications. Moreover, these plastic events redistribute stress non-locally, as if the foam were an elastic medium subjected to a quadrupolar deformation. Our findings clarify how the macroscopic elastoplastic behavior of amorphous materials emerges from the spatiotemporal stress variations induced by microstructural rearrangements.**

## 1 Introduction

Gels, pastes, emulsions, foams and colloidal suspensions are typical examples of soft amorphous materials that we commonly use in our daily lives for their unique mechanical properties, exhibiting both solid-like and liquid-like behavior. This dual nature is indeed at the root of their use in numerous applications and industrial processes, spanning sectors such as food, cosmetics, pharmaceuticals, building materials, resource extraction, and environmental remediation[1–3]. Above a critical stress, these disordered materials undergo a yielding transition from a solid to a liquid state due to irreversible topological rearrangements at the microscopic scale of their jammed constituents (colloids, grains, droplets, bubbles …). Despite their widespread use with numerous commercial products and extensive research efforts, a fundamental understanding and control of this unjamming transition remains a formidable challenge today[4,5]. This difficulty arises from the inherent complexity in directly observing and characterizing the out-of-equilibrium evolution of the network topology of the material constituents. Indeed, standard rheological tools provide only a global macroscopic mechanical response of the probed material. Thus, they lack the spatio-temporal resolution necessary to capture the heterogeneous mechanical flows of those soft disordered materials, characterized by correlated structural rearrangements that lead to slips, shear bands, or fractures[6–8].

In recent years, significant progress has been reported[9] thanks to the use of different optical techniques such as confocal microscopy[10–12], light scattering[13], or differential dynamic microscopy[14,15] coupled to rheometry, allowing to observe the microscopic dynamics of soft jammed materials, and particle rearrangements occurring during shear tests[16–18]. However, these methods present strong limitations. In particular, confocal microscopy requires the use of (fluorescent) colloidal particles - thus most of current experiments concern only colloidal glasses, while Fourier-based techniques do not provide any direct information on individual particles. More importantly, none of these techniques can offer a measure of the stress at the local scale of the jammed constituents of the materials.

Interestingly, liquid foams, archetype of soft jammed materials[19], possess remarkable properties that make them ideal candidates for characterizing stress down to the scale of their inner structure, specifically the bubbles[20]. First of all, the difference in density between the gaseous and liquid phases can be exploited as an intrinsic contrast agent for 3D imaging technique such as X-ray tomography[21–25]. New developments in X-ray imaging allowed recently to probe foam dynamics below the time resolution of one second, with notably a 3D image (a tomogram composed of thousands of projections) acquired in less than 0.5 seconds[26]. Second, the stress at the local scale of a single bubble can be inferred from the bubble's shape through the intermediate of surface tension, which was already exploited in simulations[27,28], but never in 3D experiments. Furthermore, modifying the physico-chemistry of the foaming solution and the nature of the bubble gas, one can obtain ultra-stable foams. In these foams, the ageing processes—such as liquid drainage,

---

[a] *Division of Solid Mechanics, LTH, Lund University, Lund, Sweden. E-mail: florian.schott@solid.lth.se*
[b] *Université Grenoble Alpes, CNRS, LIPhy, 38000 Grenoble, France*
[c] *Laboratoire de Physique, UMR CNRS 5672, ENS de Lyon, Université de Lyon, Lyon, France*
[d] *Swiss Light Source, Paul Scherrer Institut, 5232 Villigen, Switzerland*
[e] *Université Côte d'Azur, CNRS, INPHYNI, France*
[f] *Institut Universitaire de France (IUF), France*
[g] *Department of Physics, Technical University of Denmark DK-2800 Kgs. Lyngby, Denmark*

bubble coalescence, and coarsening—can be significantly slowed down[29], and the duration of a topological rearrangement can be controlled over a large range of time scales[30].

Therefore, we propose here an original experimental study on stable liquid foams, where we combine simultaneous time-resolved fast X-ray micro-tomography with continuous rheological shear testing in a plate-plate geometry. Our tomo-rheoscopy allows to overcome current state-of-the-art limitations (low spatial and temporal resolution, reduced field of view, turbidity/opacity of the material) to directly observe and monitor the local deformation, stress, and contact topology of each bubble within the sheared foams. First, by analyzing the shape of each segmented bubble (up to $10^5$ in a single tomogram) within the sheared foam, we obtain the deviatoric stress tensor at the scale of each individual bubble. Computing the corresponding stress field over larger volumes up to the entire foam sample leads to a measurement of the shear stress applied on the rheometer plates, which is in quantitative agreement with the independently measured torque provided by the rheometer, thus validating our local stress measurements. Then, by tracking the contact network topology of each bubble, we are able to unveil the detailed spatio-temporal mechanical response of the sheared foam at the bubble scale. We focus on elementary topological rearrangements, known as T1 events[31], which consist of the loss of contact between two bubbles and the simultaneous creation of a new contact between two neighbouring bubbles. Such neighbour-swapping events involving four bubbles represent the elementary plastic process for foams. Our tomo-rheoscopy of liquid foams allows us to reveal a universal scaling behavior in local stress build-up and relaxation associated with such elementary plastic events, from which the macroscopic visco-elasto-plastic behaviour of the foam emerges. Indeed, for the various foams probed, with different liquid fractions, and subjected to different local shear rates, we obtain a master curve describing the local stress as a function of the local strain. On average, each bubble involved in a T1 event exhibits a stress increase before the T1, with bubbles losing and gaining contact showing increases of 20% and 10% above the yield stress, respectively. This is followed by a stress drop of 40% below the yield stress after the T1 in both cases. Furthermore, we also demonstrate that these elementary plastic events spatially redistribute stress in a non-local manner, as if the foam were an elastic medium subjected to a quadrupolar deformation, in accordance with Eshelby's prediction[32].

Our original experimental setup and analysis demonstrate that the fast tomo-rheoscopy of a model soft-glassy material can offer a complete and comprehensive understanding of the yielding of the material's constitutive structural elements within a shear flow. This breakthrough opens up an entirely new avenue for research, calling for new experimental campaigns to study various types of liquid foam samples and, more broadly, other soft jammed materials submitted to different flow conditions, such as oscillatory or creep tests to uncover the fundamental causes and origin of their macroscopic rheological properties.

## 2 Results

Eight liquid foam samples were prepared as detailed in Methods. They are almost monodisperse, with an average radius varying across samples between 50 and 80 $\mu$m, and a liquid fraction between 5% and 21% constant in space and time within a few percents (see Supplementary Information). Their stability was optimised such that no significant ageing could be measured during the experimental runs.

In order to perform simultaneously continuous shear testing and time-resolved X-ray micro-tomography imaging of such liquid foam samples, we have designed and developed an innovative set-up based on a dual-motor rheometer prototype, described in Methods. We use here a plate-plate geometry of radius $R = 2.5$ mm and gap $h = 1.5$ mm, with rough plates ensuring no-slip boundary conditions (Fig. 1a). The plates are in relative rotation at a low angular speed $\Omega = 2\pi \times 10^{-3}$ rad/s. The plate-plate geometry allows for the application of different local shear rates. The local shear rate component in the $\theta z$ plane $\dot{\gamma}(r) = \Omega r/h$ increases linearly with the radial distance $r$ from the rotation axis, considering a cylindrical coordinate system $(r, \theta, z)$ centered on the rotation axis, with $z = 0$ corresponding to the lower plate. We verified that the flow field follows closely the applied shear rate (Fig. 5a): no shear banding is present in our experiments.

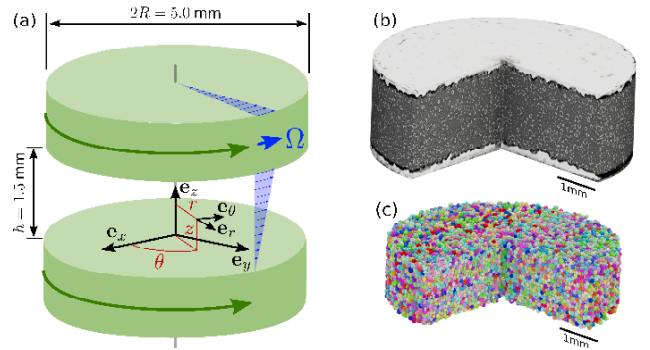

Figure 1: **Tomo-rheoscopy and image analysis** – (a) Sketch of the co-rotation rheometer geometry allowing simultaneous continuous shear testing and micro-tomography imaging. (b) Typical 3D raw image with a quarter vertical cut showing the bubbles within the gap. (c) Corresponding bubble segmented image. The bubbles are distinguished by their color.

Three-dimensional (3D) images (called tomograms) are acquired every 3 s using the fast X-ray tomography facility of the Swiss Light Source. Each experiments counts 160 tomograms, imaging the whole liquid foam sample over a volume of around 30 mm$^3$ with $2016 \times 2016 \times 800$ voxels and a voxel size equal to 2.48 $\mu$m. A typical tomogram is shown in Fig. 1b. Movies of typical experiments in various cross-sections are also provided in Supplementary Information. Bubbles are segmented as detailed in Methods, as shown in Fig. 1c.

Our experiments are run as follows. The liquid foam is placed within the rheometer gap, using a syringe. In order to reduce subsequent residual bubble deformation, before each test, we first apply a pre-shear with a high relative rotation speed of 50 $\Omega$ for 20 s in one direction followed by a shear in the opposite direction at the same rate for an equal duration. The sample is let to relax for 80 s, and then sheared at $\Omega$ while simultaneously imaged. Our shear experiments last 480 s. This procedure yields a unique and unprecedented set of experimental data, with an extensive statistics, giving access to the position, shape, contact topology of each bubbles within sheared liquid foams, while measuring the temporal evolution of the global shear stress by the rheometer[26]. As we will explain in the next section, based on the analysis of the shape of the bubble surface, one can moreover obtain a measurement of the stress at the local bubble scale.

## 2.1 Multi-scale stress measurements – *from the bubble to the foam scale*

At low enough shear rates, viscous stresses are negligible, and the only contribution to the stress is associated with the interfaces through the surface tension $\gamma$. The non-isotropic part of the stress is given by $\sigma_{ij} = \frac{\gamma}{V} \int (\frac{1}{3}\delta_{ij} - n_i n_j) \, \mathrm{d}S$[28,33,34], where integration is carried on all interface elements $\mathrm{d}S$, with unit normal vector **n** of components $n_i$, contained in a volume $V$; $\delta_{ij}$ is the Kronecker unit tensor, $\delta_{ij} = 1$ if $i = j$ and 0 otherwise. Therefore, meshing the interface of each labelled bubble of our sheared foams, with triangular surface elements, as illustrated in Fig. 2a, we could compute this stress tensor at the bubble scale of volume $V$, by performing an integration over all those triangular surface elements $\mathrm{d}S$, of normal unit vector **n**.

Then, the macroscopic stress in larger volumes containing numerous bubbles is obtained by averaging the stress of individual bubbles whose centroids lie in such volumes. It is expressed in cylindrical coordinates inside tori to account for the axisymmetric geometry. One torus is defined by the volume containing all the bubbles between $r$ and $r + \Delta r$. In practice, the radial axis is divided into 10 sections that typically contain 84-345 and 1970-6987 bubbles for the smallest and largest radii, for the largest and smallest bubble size, respectively.

We first consider this upscaling procedure for a shear experiment of a dry foam, with a liquid fraction of $\phi_\ell = 8\%$, composed of bubbles of average radius of $52\mu$m (series 2 in Table 1). As expected with the geometry of our set-up, the relevant and main stress component is $\sigma_{\theta z}$ during our shear experiments, even though one can also notice the emergence of non-negligible normal stress components $\sigma_{\theta\theta}$ and $\sigma_{zz}$ (see the Supplementary Information, for the various components of the stress tensor). We plot the time evolution of the stress component $\sigma_{\theta z}$ at various distances $r$ from the rotation axis in the inset of Fig. 2b. This stress component first increases before reaching a plateau. The initial increase is sharper for larger values of $r$, since the local shear rate component in the $\theta z$ plane $\dot\gamma(r) = \Omega r/h$ increases linearly with $r$. Accordingly, plotting $\sigma_{\theta z}$ as a function of the local amplitude of the applied deformation $\gamma = \dot\gamma(r)t$ rescales extremely well all our data (Fig. 2b). The trend of this master curve is classical for a yield stress fluid. Specifically, all data follow a unique elastic loading (with a slope giving the elastic modulus $G$) until a steady-state strain $\gamma_{ss}$, above which the stress saturates at a steady-state shear stress value $\sigma_{ss}$. The precision of our measurements reveals a slight effect of the local shear rate $\dot\gamma$ on this stress amplitude $\sigma_{ss}$: the larger the shear rate, the larger the steady-state shear stress; nevertheless, these variations are minimal. At most, there is a 20% increase of the steady-state shear stress when the shear rate is tripled, indicating that the liquid foam experiences a quasi-static flow. One can also notice that this shear stress value $\sigma_{ss}$ exhibits a slight drift, which can be attributed to a small yet measurable decrease in the foam's liquid fraction over the course of the experiment (as shown in Supplementary Information). Nevertheless, such data obtained from our local stress measurements give access to the macroscopic intrinsic rheological properties of our liquid foams, such as the elastic modulus $G$, the yield stress $\sigma_Y$ and the yield strain $\gamma_Y = \sigma_Y/G$. Indeed, the yield stress $\sigma_Y$ can be obtained through an affine extrapolation of the steady-state shear stress $\sigma_{ss}$ in the limit of shear rates approaching zero, $\dot\gamma \to 0$ (see Methods). We show notably that our measurements are in quantitative agreement with those obtained from classical rheological measures[35,36], and notably their evolution with the liquid fraction.

Finally, we can compute the macroscopic stress at the foam scale (based on our fast imaging and detailed analysis at the local bubble interface scale) and compare such measurement to the independently measured torque $\tau^{\mathrm{rheometer}}$ provided by the rheometer. Indeed, integrating the local stress exerted by the foam on the rotating plate yields the following prediction: $\tau^{\mathrm{image}} = 2\pi \int_0^R r^2 \sigma_{\theta z}|_{z=h} \, \mathrm{d}r$. We plot the time evolution of $\tau^{\mathrm{rheometer}}$ and $\tau^{\mathrm{image}}$ for foams of different liquid fractions in Fig. 2c and d. Remarkably, our global torque estimation based on the analysis of the shape of the bubbles' surfaces is in excellent quantitative agreement with the independent measurement given by the rheometer, for our various shear experiments performed over different liquid foams, for which we systematically varied their liquid fraction.

Therefore, we demonstrate here our ability to measure and monitor the stress from the local bubble scale up to the global foam scale, during a quasi-static shear test. We can now take advantage of such local stress measurement to study the plastic properties of our liquid foams.

## 2.2 Universal mechanics of an elementary topological rearrangement

We can track and monitor the temporal evolution of the contact network topology of each bubble in the sheared liquid foams. More specifically, we can detect topological changes corresponding to either the loss or creation of contacts between bubbles. Here, we propose to focus solely on the so-called T1 events, the elementary topological rearrangements involving the simultaneous loss of a contact between two bubbles and the creation of a new contact between two other neighbouring bubbles. For a given radial distance, which corresponds to a given shear rate, we recorded between 5000 and 33000 T1 events for the smallest and largest radii, respectively, considering only the events occurring on the plastic steady-state shear stress plateau.

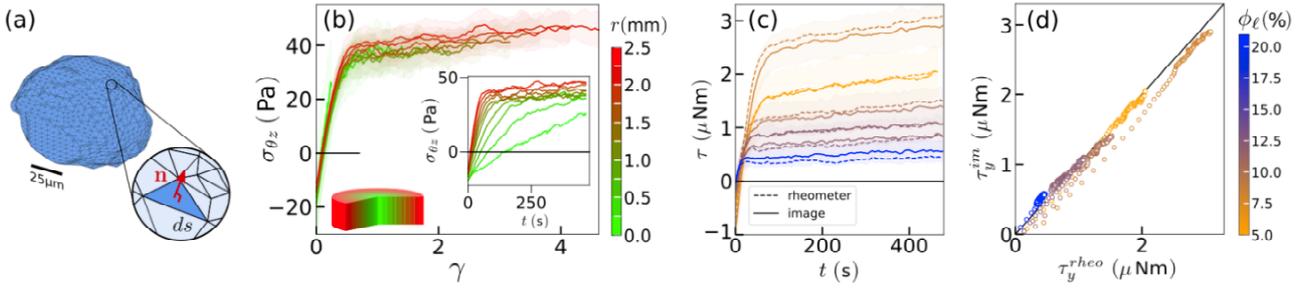

Figure 2: **Multi-scale stress measurements** –
(a) Example of a meshed bubble interface over which the Batchelor stress tensor is integrated with each individual triangle surface $ds$ and normal unit vector $\mathbf{n}$. (b) The shear stress $\sigma_{\theta z}$ at various radius rescales when shown as a function of the local deformation $\gamma$. $\sigma_{\theta z}$ as a function of time is shown in the sub-figure. (c) Rheometer and local torque measures as a function of time for various series. The sub-figure shows the local torque as a function of the rheometer torque on the yield plateau (from 100 s to 480 s).

We plot in Fig. 3a and b the averaged shear stress component $\sigma_{\theta z}$ of the bubbles losing a contact, as a function of time $\Delta t$, with the convention that $\Delta t = 0$ marks the occurrence of the topological change. The time evolution of this quantity $\sigma_{\theta z}$ is studied as a function of the local shear rate (corresponding to different radial distances $r$) for a given liquid fraction (Fig. 3a) or as a function of the liquid fraction for a given local shear rate (Fig. 3b and c). Interestingly, our data extracted from very different experimental conditions display the same behaviour: far from the T1 event, the averaged shear stress component $\sigma_{\theta z}$ takes a value compatible with the plastic steady-state shear stress plateau $\sigma_{ss}$, previously quantified for all bubbles at the foam scale (Fig. 2b). Then, $\sigma_{\theta z}$ increases until it reaches a maximum just before the T1 event, when it sharply decreases to a minimum value, lower than the plastic plateau, marking the post-T1 relaxation. It then increases anew before recovering the value of the plastic plateau. One can notice that the larger the shear rate or the liquid fraction, the faster the variations associated with the imposed deformation. The stress relaxation duration $\Delta t_{T1}$ associated with a T1 event (measured as the time difference when reaching the corresponding maximum and minimum shear stress values) varies from 7 to 12 s, depending on the liquid fraction of the probed foam (Fig. 3f). Such a time scale is consistent with measurements performed over a sheared cluster of four bubbles, formulated with similar rigid surfactants solutions, and shown to be controlled by the ratio between surface tension and interfacial viscous forces[30]. This characteristic time is significantly shorter than the timescale $1/\dot{\gamma}$ of the imposed deformation, making our shear experiments effectively quasi-static.

Remarkably, all data collapse onto a master curve when $(\sigma_{\theta z} - \sigma_{ss})/\sigma_{ss}$ is plotted as a function of $\Delta\gamma/\gamma_{ss}$ (Fig. 3d), with $\Delta\gamma = \dot{\gamma}\Delta t$ and $\gamma_{ss} = \sigma_{ss}/G$. This data collapse (including measurements for various shear rates and liquid fractions) is obtained thanks to the use of the macroscopic foams properties $\sigma_{ss}$ and $\gamma_{ss}$. Such a result highlights that the macroscopic visco-elasto-plastic behaviour of our 3D liquid foams reflects the local mechanical properties of the individual bubbles, and specifically the local stress variations associated with their structural rearrangements. Quantitatively, variations occur within a range of $\Delta\gamma = \pm\gamma_{ss}$, where the maximum stress represents a 20% increase with respect to the steady-state shear stress value $\sigma_{ss}$, followed by a subsequent relative stress drop of 40%.

Strikingly, the same phenomenology and a similar data collapse were obtained when analysing the stress variations for the two bubbles gaining contacts during the same T1 events. Nevertheless, in this case, the stress increase is only 10%, marking a difference that could be exploited to distinguish and sort the behaviour of those yielding bubbles. The post-T1 relaxation for the bubbles gaining a contact maintains the same relative stress drop amplitude of 40% (Fig. 3e).

In both cases, these variations highlight the intimate behaviour of the jammed bubbles which need first to deform before they can relax stress. The smooth continuous aspect of this dimensionless local stress-strain function can be attributed to the disordered structure of the bubble arrangement within our foams, which indeed appears very different from the sharp discontinuous functions obtained in simulations for ordered and crystalline bubble assemblies[37,38]. As such, this fine characterisation of the mechanical behaviour of the elementary structural elements of our soft jammed materials, (here the bubbles of a liquid foam) constitutes an important result, that will be particularly useful as a raw input in numerical simulations, and more generally for validating models of the yielding of liquid foams and other amorphous materials[39,40].

## 2.3 Non-local quadrupolar stress redistribution of a T1 event

We can now extend our analysis further to investigate the stress variations that occur during such elementary topological rearrangements in their spatial surroundings (Fig. 4a).

First of all, a given T1 event is characterized by the directions of the lost and new contacts, which are represented by the unit vectors **a** and **b**′, respectively. These vectors are oriented along the lines connecting the centres of the corresponding bubbles (Fig. 4b). The orientation distributions of those two vectors, shown in Supplementary Information, display peaks in the $\theta z$ plane at directions close to $\pm 45°$ to the azimuthal and axial directions. These orientations correspond to the principal directions of the strain rate tensor in simple shear geometry, indicating that T1 events relax the elastic energy accumulated due to the imposed deformation.

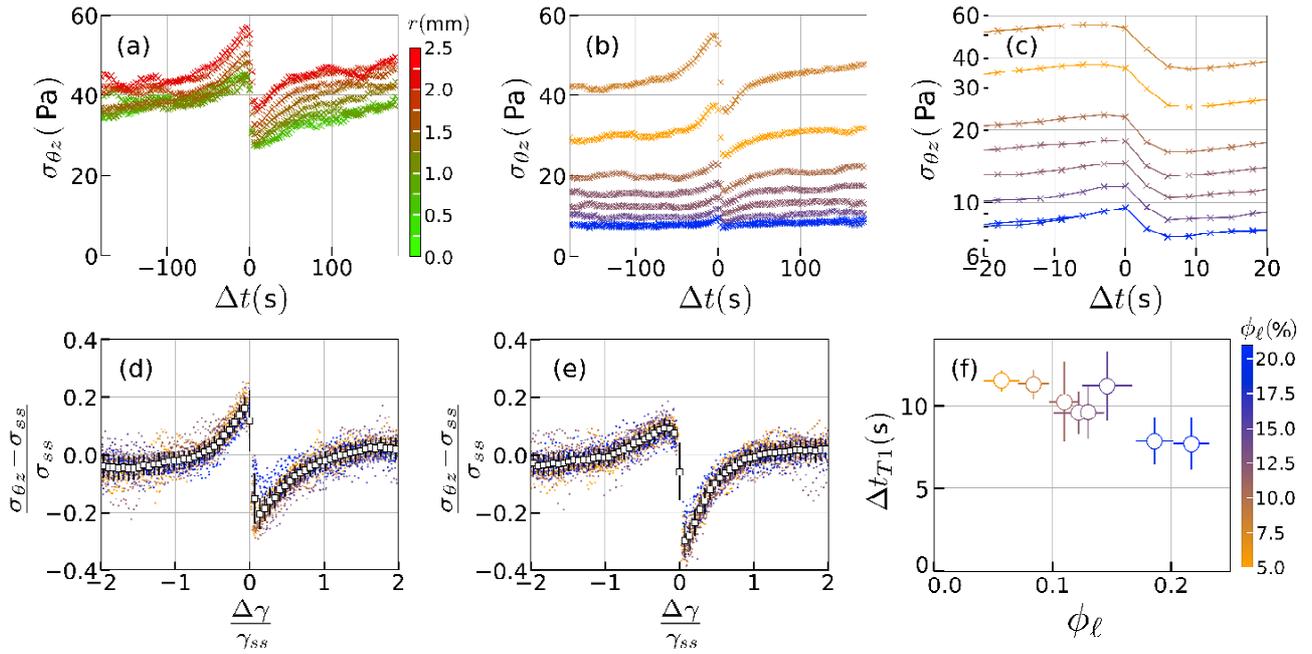

Figure 3: **Universal mechanical signature of a T1 event** –
(a) Time evolution of the averaged shear component $\sigma_{\theta z}$ for bubbles losing a contact during the T1 event at different shear rates in the range 0.003 - 0.010 s$^{-1}$, corresponding to different radial distances from the rotation axis within the rheometer gap. (b) The same approach for experiments with different liquid fractions and a fixed shear rate of 0.009 s$^{-1}$, corresponding to the radial position $r \in [2.0, 2.25]$. (c) Same plot as (b) with a close-up between -20 and 20 s to focus on the stress relaxation and a log scale for the stress axis. (d) Collapse of the $(\sigma_{\theta z} - \sigma_{ss})/\sigma_{ss}$ as a function of $\gamma/\gamma_{ss}$. (e) Collapse of the same quantity for bubbles gaining contact during the same T1 event. (f) Stress relaxation duration $\Delta t_{T1}$, as measured in (c), as a function of the liquid fraction $\phi_\ell$. Error bars are obtained from statistics over all events.

For each T1 event, we define a local frame of reference $\{O, \mathbf{a}, \mathbf{b}, \mathbf{c}\}$ with $O$ marking the centroid of the four bubble centers involved in the T1, and $(\mathbf{a}, \mathbf{b}, \mathbf{c})$ as the basis vectors defined as follows: **a** is the unit vector in the direction of the lost contact, as defined previously. $\mathbf{b} = [\mathbf{b}' - (\mathbf{a} \cdot \mathbf{b}')\mathbf{a}]/(1 - \mathbf{a} \cdot \mathbf{b}')$ is the unit vector in the orthogonal direction to the lost contact in the plane $(\mathbf{a}, \mathbf{b}')$ (**b** is introduced because the directions of the lost contact **a** and of the new contact **b**′ are never perfectly orthogonal). $\mathbf{c} = \mathbf{a} \times \mathbf{b}$ is the unit vector in the direction orthogonal to the plane of the lost and new contacts. In this local frame, we can obtain a spatial map of the temporal stress variations occurring around a T1.

In practice, we compute the differences of the various components of the stress tensor $\Delta \sigma_{ij} = \sigma_{ij}(t+1) - \sigma_{ij}(t)$ (with $i$ and $j$ equal to $a$, $b$ or $c$), for two successive images during which a T1 event occurs, for each bubble whose centroid is located at a distance up to 8 bubble radius away from the T1 centre. We can then perform an average of those stress variations within 3D cubical zones (with a resolution of 25 x 25 x 25 voxels), measured over all the T1 detected at a given radial position (corresponding to a given local shear rate).

We consider our reference experiment corresponding to a dry foam (series 2 in Table 1), and include all the T1 events detected across all radii and times. The analysis is restricted to T1 events where all four bubble layers surrounding the T1 are fully captured, discarding events centered within the three bubble layers closest to the lower and upper plates. This filtering reduces the dataset to 13300 events out of the original 33000. The average steady-state shear stress, $\sigma_{ss} \simeq 40$ Pa, is used as the typical stress scale. Figures 4c, d and e display cross-sections the 3D spatial maps of the stress redistribution around such elementary topological rearrangements. First of all, one can observe that a T1 event has a strong, non-local impact in its surroundings, with a non-trivial spatial perturbation of the stress field and specific angular dependencies in various directions. Indeed, the variation of the shear stress component $\Delta \sigma_{ab}$ displays a very clear quadrupolar pattern with increases and decreases alternating every eighth of a turn in the $(a, b)$ plane. Given the orientation of (**a**,**b**), the poles of the quadrupolar shape are oriented at approximately $\pm 45°$ in the $(\theta, z)$ plane. On the other hand, the stress variation component $\Delta \sigma_{aa}$ tends to decrease, and the component $\Delta \sigma_{bb}$ to increase, along the lost contact direction $a$ and new contact direction $b$, consistently with the main displacement of the corresponding bubbles across the T1. While the quadrupolar pattern was already reported in 2D[41,42], the full 3D picture appears much richer than the simple 2D case: for instance, the $cc$ component tends to increase in the $a$ direction, but to decrease in the $b$ direction. This means that the bubbles losing contact, respectively gaining contact, tend to relax by becoming less prolate, respectively less oblate. We furthermore observe clear angular patterns emerging in the two planes $ac$ and $bc$.

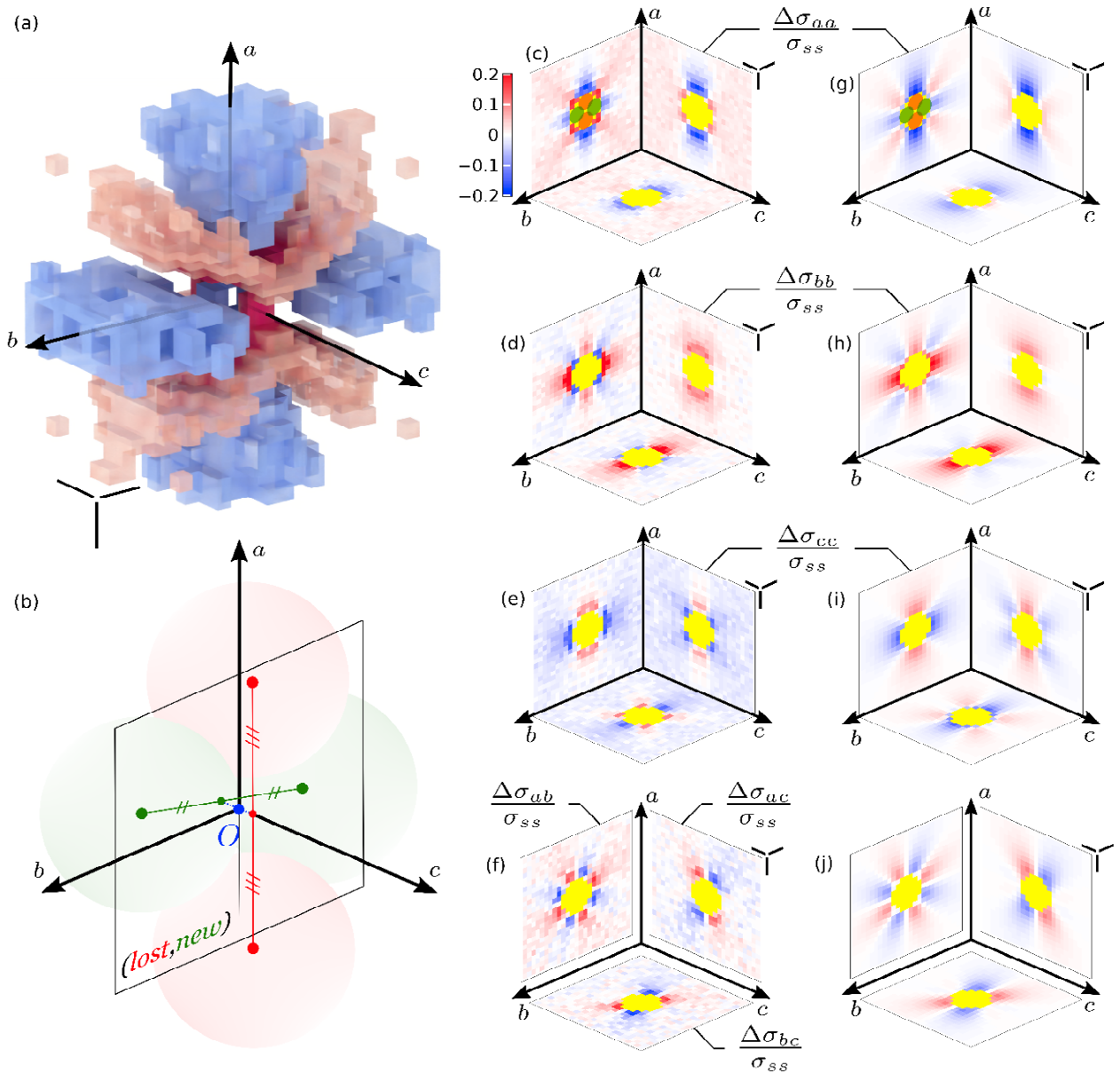

Figure 4: **Non-local quadrupolar stress redistribution of a T1 event** –
(a) 3D representation of the average stress variation in the T1 frame of reference, using the example of the $\Delta\sigma_{aa}$ component, displaying a quadrupolar shape. The positive and negative deviations are displayed in red and blue, respectively, with transparency. (b) Definition of the T1 frame of reference $\{O, \mathbf{a}, \mathbf{b}, \mathbf{c}\}$ defined from the centers of the four bubbles involved in the T1 event. Bubbles losing and gaining contact are shown in green and orange, respectively. (c-e) Average stress variations after the T1 event in the mid-planes $(a,b)$, $(a,c)$ and $(b,c)$. Only non-zero cross-sections are shown. (d-h) Average stress variations obtained using the model with the parameter $\epsilon^s$ set to 0.023. In (c) and (g), the average positions and shapes of the four bubbles involved in the T1 events are depicted with the colors defined in (b). For comparison between experiments and model, we superimposed a spherical mask, in yellow, that marks the region encompassing the four bubbles involved in the T1 events. In all panels except (b), the three scale bars represent the mean bubble diameter in the three directions.

To go further, we can compare our measurements of the stress variations in the spatial surrounding of a T1 event to the ones obtained for an elastic medium due to an imposed quadrupolar deformation, $\varepsilon^s = \varepsilon^s(\mathbf{a}\otimes\mathbf{a} - \mathbf{b}\otimes\mathbf{b})$. Indeed, such a deformation corresponds to the fact that the two bubbles losing contact get further apart while two bubbles gaining contact get closer. Thus, we assimilate the effect of a T1 as an imposed displacement $\mathbf{u}^s = \varepsilon^s \cdot \mathbf{x}$ at the surface of a sphere of radius $a$ centred on the T1. Considering that the foam reacts to this applied strain as an elastic Hookean material of Young's modulus $E$ and Poisson ratio $\nu = 1/2$ (the foam being quasi-incompressible), we can compare our measurements to Eshelby's prediction who solved this mechanical problem in a seminal work[32]. We provide those predicted stress components in Supplementary Information. They share the common form $\Delta\sigma_{ij} = E\varepsilon^s f_{ij}(\mathbf{x}/a; \nu)$, where the dimensionless functions $f_{ij}$ quantify their spatial dependencies; notably, at large distances $x \gg a$, they decay as $1/x^3$. We use $\varepsilon^s$ as a single fitting parameter, for those various components $\Delta\sigma_{ij}$ to challenge such prediction. The agreement with our data is remarkable, including all the involved angular dependencies (see Figs. 4f, g and h for cross-sections of the 3D maps, and Supplementary Information for quantitative comparisons at specific positions). Furthermore, we could check that Eshelby's prediction is also verified for our various experimental parameters, *e.g.* at different radial position $r$ of the T1 and thus for different local shear rates $\dot{\gamma}$, and the various foams' liquid fractions $\phi_\ell$. While $\varepsilon^s$ appears independent of the local shear rates $\dot{\gamma}$, it decreases systematically with the liquid fraction (see Supplementary Information). This value should be correlated to the foam yield strain $\gamma_Y$ since we are computing the variations of the stress field over two successive images, with a time scale of 3 seconds, while a T1 rearrangement can last up to more than 10 seconds. For instance, for our reference experiment, we measure $\varepsilon^s = 0.068 \pm 0.002$. Considering the cumulative deformation over the duration of a T1 for such dry foam ($\sim 12$ s), gives indeed a value close to $\gamma_Y \sim 0.23$.

Overall, this constitutes the first experimental proof of the validity of the elastoplastic approach[39] for an amorphous material in 3D, and directly on the local stress.

imposed local shear rates. Furthermore, we demonstrate that such T1 plastic events redistribute stress non-locally and anisotropically, imposing quadrupolar deformations, within the foam behaving as an elastic medium.

While our results call for new experiments and will likely inspire studies from other research groups, it is important to note that we have focused here on the elementary T1 events. These events correspond to the loss of contact between two bubbles, simultaneously compensated by the creation of a new contact between two other neighboring bubbles. However, such events account for only 25% of the various plastic events detected in our experiments on plastic flows, with the remaining events involving changes in the network topology that affect a larger number of bubbles. As such, we plan to analyze the statistical properties of these clusters of plastic rearrangements, including their size, shape, orientation, localization, and duration. More generally, future work will investigate the spatiotemporal correlations of the corresponding local stress drops and bubble displacements, which may lead to avalanches and eventually to the formation of shear bands. It is important to notice that indeed the local stress build-up before triggering a T1 is about 20% with respect to the steady-state shear stress value $\sigma_{ss}$, and thus $\sigma_Y$ in the limit $\dot{\gamma} \to 0$. Therefore, such a perturbation is far to be negligible, and as such, we can expect to detect subsequent "aftershocks", triggered plastic events. Nevertheless, a detailed statistical analysis will require to investigate a wider range of applied shear rates to reach globally heterogeneous flows. The latest improvement in fast X-ray micro-tomography will be invaluable (one should remind actually that the temporal resolution gained several orders of magnitude over the last decades - in 2005 the acquisition time for a single tomogram was about 150 s, for spatial resolution with a voxel size around 10 $\mu$m)[43].

## 3 Discussion

To conclude, we have demonstrated our ability to directly observe and monitor the 3D local displacements, stresses, and contact topology of the elementary microstructural components of a model soft glassy material under quasi-static shear flow. We could achieve such a challenging feat, thanks to the development of an innovative setup that leverages state-of-the-art 3D imaging, using time-resolved fast X-ray micro-tomography, combined with shear rheometry of well-chosen ultra-stable liquid foams. Through an accurate image analysis, our tomo-rheoscopy technique enables multi-scale stress measurements allowing us to establish links between the building blocks of these soft jammed materials (the bubbles) and the macroscopic intrinsic properties of the foams characterising their yielding behaviour. Indeed, we unveil the *universal* mechanical signature of the elementary topological rearrangement (involving the neighbour swapping of four bubbles), with a local stress-strain function independent of the foam's liquid fraction and the

# 4 Methods

**Foam generation.** The foaming solution was prepared by following the protocol of Golemanov et al.[44]. First, 6.6% sodium lauryl ether sulfate (SLES) and 3.4% cocamidopropyl betaine (CAPB) in mass were mixed in ultra-pure water; then 0.4% of myristic acid (MAc) was dissolved into the solution by stirring and heating at $60^o$C for one hour; this solution was finally diluted 20 times with a glycerol/water mixture with a mass ratio of 50/50 to obtain the foaming solution. Foams were generated within a co-flow microfluidic setup, by simultaneously introducing a foaming solution and perfluorohexane-saturated air. Precise control of solution flow rate and gas pressure was achieved using a Harvard Apparatus PHD Ultra syringe pump and a Fluigent MFCS-FLEX pressure controller, respectively. Subsequently, selected foams were subjected to centrifugation using an Eppendorf 5702 centrifuge at different rotation rates, to decrease their liquid fraction. In this paper, we focused on eight different foam samples, for which the liquid fraction $\phi_\ell$ and bubble size distributions (characterized through the Sauter mean radius $R_{32}$) were directly measured in situ from tomographic images (see the corresponding subsection *Image processing*) and reported in Table 1.

| Series | $\phi_l$ (%) | $R_{32}$ ($\mu$m) |
|---|---|---|
| 1 | 5.7 | 78.4 ± 3.9 |
| 2 | 8.4 | 52.1 ± 0.9 |
| 3 | 11.0 | 61.0 ± 0.8 |
| 4 | 12.2 | 76.3 ± 3.6 |
| 5 | 13.0 | 61.3 ± 1.3 |
| 6 | 14.6 | 77.3 ± 4.3 |
| 7 | 18.6 | 61.3 ± 0.5 |
| 8 | 21.7 | 48.7 ± 0.6 |

Table 1: Liquid fraction and average bubble radius for the eight series.

**Co-rotation rheometer prototype.** A dual-motor rheometer (Anton Paar MCR 702 TwinDrive) was adapted for co-rotation, enabling precise application of controlled macroscopic deformation, torque measurement, and simultaneous 3D tomographic image acquisition. This innovative prototype was a collaborative effort between PSI and the Laboratory of Food Process Engineering at ETH Zürich. The 2.5 mm radius plates were 3D printed and covered with sandpaper (P80). The foam was placed inside the rheometer, using a syringe. The gap $h$ was reduced to 1.5 mm, and the excess foam was removed from the outer free edge. The sample size corresponded to the available tomography field of view, allowing to image the whole sample while it is sheared.

**Tomography imaging.** Imaging was conducted at the TOMCAT beamline, Paul Scherrer Institute, Switzerland, utilizing a monochromatic X-ray beam with a photon energy of 16 keV. For each time series, high resolution angular projections at 0.5 ms exposure time while rotating the sample at 1 Hz were recorded. Images were acquired with a dedicated fast X-ray detector system[45] placed 25 cm downstream the sample. One 3D image reconstruction was captured every 3 s, covering a volume of 1.5×5×5 mm$^3$ with a voxel size equal to 2.48 $\mu$m. Before the tomographic reconstruction the projected thickness map[46] were retrieved in the phase contrast radiographic projections.

**Shear test protocol** To obtain comparable deformation history, each of the eight foam samples considered in this study were first sheared with a relative rotation speed of 50 mHz for 20 s in one direction followed by an equal duration in the opposite direction. After 80 s at 0 Hz, the foam was sheared at 1 mHz and simultaneously imaged for 480 s. Given the plate-plate geometry, the shear rate is a linearly increasing function of the radial position within the rheometer. The maximum value is reached at the periphery, with a value of 0.01 s$^{-1}$.

**Image processing.** The 3D images were phase-segmented with an Otsu threshold. The binary values (0 and 1) correspond to the gas and liquid phases respectively. The images were then binned by joining eight neighbouring voxels and each individual bubble volume was segmented with the ITK watershed[47]. This step is crucial to correctly reconstruct the films between bubbles, which are too thin to be directly detected on the binary images. Physical properties at the scale of the foam were extracted and reported in the Table 1. For each bubble $i$, we obtain the volume $V_i$ and the equivalent radius $R_i = (3V_i/4\pi)^{1/3}$. From this, we obtain the average bubble radius for a given series as $R_{32} = \langle R_i^3 \rangle / \langle R_i^2 \rangle$, where the average is taken over all bubbles. Notice that our foams are almost monodisperse with standard deviation of the bubble radius distribution at least one order of magnitude lower than the average. We also checked that the distribution of bubble sizes remained spatially uniform throughout each experiment's duration. The liquid fraction $\phi_l = n_{liquid}/(n_{liquid} + n_{gas})$ is defined as the ratio of the number of voxels in the liquid phase to the total number of voxels in a phase segmented image, with $n_{liquid}$ and $n_{gas}$ the number of voxels in the liquid and gas phases respectively. The bubble volumes were extracted from the bubble segmented images.

**Local velocity field measurements** The displacement of bubbles between two successive images is obtained by Discrete Digital Volume Correlation[47]. Bubble displacements are then averaged to infer velocity fields. An example of velocity measurements, $v_\theta$, for a given series is shown in Fig. 5a, with the convention that the velocity is zero at the contact with the lower plate. Plotting $v_\theta$ as a function of $zr$ reveals a proportionality relationship, as expected from the imposed boundary conditions, $v_\theta = \frac{\Omega}{h} zr$, with $\Omega = 2\pi \times 10^{-3}$ rad.s$^{-1}$ is the relative angular speed and $h = 1.5$ mm the gap between the rheometer plates.

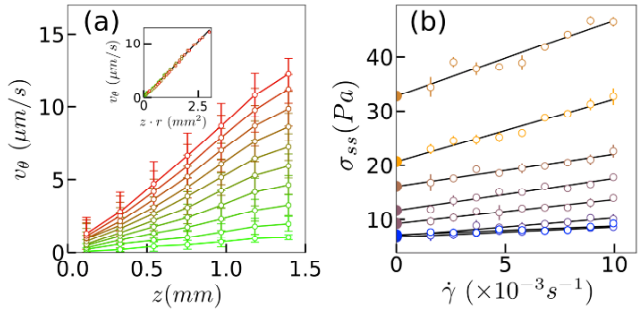

Figure 5: (a) Average velocity component $v_\theta$ as a function of height $z$ for various radii $r$ (series 2 in Table 1). Insert: $v_\theta$ as a function of $zr$. (b) Average steady-state shear stress $\sigma_{ss}$ inferred from local measurements as a function of the shear rate $\dot\gamma$ for the 8 series described in Table 1. The yield stress $\sigma_Y$ is obtained by linear extrapolation as $\dot\gamma \to 0$.

**Mechanical characterization of the liquid foams**
From the global stress-strain curves computed from our local image analysis, one can extract relevant rheological properties of our foam samples: the shear modulus $G$, the yield stress $\sigma_y$, as well as the yield strain $\gamma_y$. The yield stress is determined through a linear extrapolation of the function $\sigma_{ss}(\dot\gamma)$ as $\dot\gamma \to 0$ (Fig. 5b). The yield strain $\gamma_Y$ is then calculated using the formula $\gamma_Y = \sigma_Y/G$. As shown in Fig. 6, their evolution with the liquid fraction $\phi_\ell$ is in quantitative agreement with data from the literature obtained from classical rheological measurements[35,36].

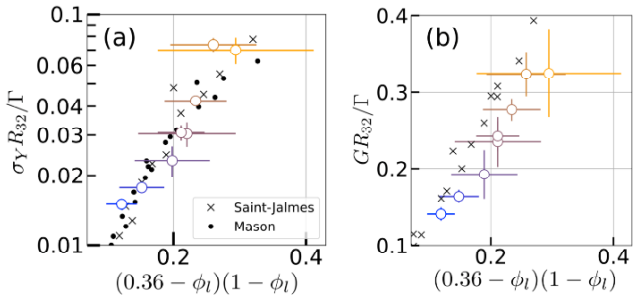

Figure 6: Dimensionless yield stress $\sigma_Y$ and shear elastic modulus $G$ as functions of the liquid fraction, as proposed by Saint-jalmes and Durian[36].

# Acknowledgements

We wish to thank the Swedish Research Council for funding this project (grant No. 2019-03742); Landshövding Per Westlings Minnesfond, Sigfrid and Walborg Nordkvists, and Knut and Alice Wallenbergs Foundation for additional traveling found (grant No. RLh2023-0011). We acknowledge the Paul Scherrer Institut, Villigen, Switzerland for provision of synchrotron radiation beamtime at the TOMCAT beamline X02DA of the SLS. The Tomo-Rheoscope used in this study was funded by the Swiss National Science Foundation (Grant No. 205311, https://data.snf.ch/grants/grant/205311). S.S. thanks support from the CNRS and ENS de Lyon, through the IRP (D-FFRACT). We thank Thibaut Divoux, Sebastien Manneville, Emanuela Del Gado, Kirsten Martens, Elisabeth Lemaire and Olga Volkova for fruitful discussions.

# Author contributions

# Additional information